\documentstyle[epsfig]{aprim10}
\input{epsf}

\newif\ifAMStwofonts

\newcommand{\beq}{\begin{equation}}
\newcommand{\eeq}{\end{equation}}
\newcommand{\beqn}{\begin{eqnarray}}
\newcommand{\eeqn}{\end{eqnarray}}

\newcommand{\AU}{{\rm AU}}

\ifoldfss
  \ifCUPmtlplainloaded \else
    \NewTextAlphabet{textbfit} {cmbxti10} {}
    \NewTextAlphabet{textbfss} {cmssbx10} {}
    \NewMathAlphabet{mathbfit} {cmbxti10} {} 
    \NewMathAlphabet{mathbfss} {cmssbx10} {} 
  \fi
  \ifAMStwofonts
    \ifCUPmtlplainloaded \else
      \NewSymbolFont{upmath} {eurm10}
      \NewSymbolFont{AMSa} {msam10}
      \NewMathSymbol{\upi}     {0}{upmath}{19}
      \NewMathSymbol{\umu}     {0}{upmath}{16}
      \NewMathSymbol{\upartial}{0}{upmath}{40}
      \NewMathSymbol{\leqslant}{3}{AMSa}{36}
      \NewMathSymbol{\geqslant}{3}{AMSa}{3E}

    \fi
  \fi
\fi 

\ifnfssone
  \newmathalphabet{\mathit}
  \addtoversion{normal}{\mathit}{cmr}{m}{it}
  \addtoversion{bold}{\mathit}{cmr}{bx}{it}
  \newmathalphabet{\mathbfit} 
  \addtoversion{normal}{\mathbfit}{cmr}{bx}{it}
  \addtoversion{bold}{\mathbfit}{cmr}{bx}{it}
  \newmathalphabet{\mathbfss} 
  \addtoversion{normal}{\mathbfss}{cmss}{bx}{n}
  \addtoversion{bold}{\mathbfss}{cmss}{bx}{n}
  \ifAMStwofonts
    \ifCUPmtlplainloaded \else
      \UseAMStwoboldmath
      \makeatletter
      \new@mathgroup\upmath@group
      \define@mathgroup\mv@normal\upmath@group{eur}{m}{n}
      \define@mathgroup\mv@bold\upmath@group{eur}{b}{n}
      \edef\UPM{\hexnumber\upmath@group}
      \new@mathgroup\amsa@group
      \define@mathgroup\mv@normal\amsa@group{msa}{m}{n}
      \define@mathgroup\mv@bold\amsa@group{msa}{m}{n}
      \edef\AMSa{\hexnumber\amsa@group}
      \makeatother
      \mathchardef\upi="0\UPM19
      \mathchardef\umu="0\UPM16
      \mathchardef\upartial="0\UPM40
      \mathchardef\leqslant="3\AMSa36
      \mathchardef\geqslant="3\AMSa3E
    \fi
  \fi
\fi 

\ifnfsstwo
  \DeclareMathAlphabet{\mathbfit}{OT1}{cmr}{bx}{it}
  \SetMathAlphabet\mathbfit{bold}{OT1}{cmr}{bx}{it}
  \DeclareMathAlphabet{\mathbfss}{OT1}{cmss}{bx}{n}
  \SetMathAlphabet\mathbfss{bold}{OT1}{cmss}{bx}{n}
  \ifAMStwofonts
    \ifCUPmtlplainloaded \else
      \DeclareSymbolFont{UPM}{U}{eur}{m}{n}
      \SetSymbolFont{UPM}{bold}{U}{eur}{b}{n}
      \DeclareSymbolFont{AMSa}{U}{msa}{m}{n}
      \DeclareMathSymbol{\upi}{0}{UPM}{"19}
      \DeclareMathSymbol{\umu}{0}{UPM}{"16}
      \DeclareMathSymbol{\upartial}{0}{UPM}{"40}
      \DeclareMathSymbol{\leqslant}{3}{AMSa}{"36}
      \DeclareMathSymbol{\geqslant}{3}{AMSa}{"3E}
    \fi
  \fi
\fi 

\ifCUPmtlplainloaded \else
  \ifAMStwofonts \else 
    \def\upi{\pi}
    \def\umu{\mu}
    \def\upartial{\partial}
  \fi
\fi

\title[The Vega Debris Disc]{
On the Vega Debris Disc's Dust Grains: 
Short-Lived or Long-Lived ?
}

\author[Jiang and Yeh]
       {Ing-Guey Jiang$^1$ and Li-Chin Yeh$^2$\\
        $^1$Department of Physics and Institute of Astronomy, 
National Tsing-Hua University, 
Hsin-Chu, Taiwan\\
        $^2$Department of Applied Mathematics,
National Hsinchu University of Education, Hsin-Chu, Taiwan 
}
\date{}

\pagerange{\pageref{firstpage}--\pageref{lastpage}}
\pubyear{2008}

\begin{document}

\maketitle
 
\label{firstpage}

\begin{abstract}
Through Spitzer Space Telescope's observations, Su et al. (2005) show 
that the Vega debris disc is dominated by grains which are small enough 
to be blown out by radiation pressure.
This implies the lifetime of Vega debris disc's grains is 
relatively short, about
1000 years, and a continuous dust production is necessary to maintain
the observed debris disc.
However, Krivov et al. (2006)'s theoretical calculations show that the
Vega debris disc is dominated by 10 $\mu$m grains, which would be
in bound orbits and thus long-lived, provided that the disc is in a 
steady state. In order to solve the above contradiction,
through dynamical simulations, we determine the grains' 
orbital evolutions and density profiles and seek a model 
of size distribution which can reproduce the observed surface brightness.
Our results show that a self-consistent dynamical model with a $1/R$ disc 
density profile
can be constructed when the grains have 
a power-law size distribution.
Moreover, both types of models, dominated by short-lived and 
long-lived grains, are consistent with the observational data. 

\end{abstract}


\section{Introduction}

It is still unclear how the debris discs form.
In the standard scenario, debris discs are constructed at the time when 
planetesimals are forming and colliding frequently. Thus, 
debris discs can be generated only when there are km-sized planetesimals 
colliding and producing huge amount of new dust grains.
This would take place at the stellar age of million years
when the original seed grains grow to become
km-sized planetesimals (Cuzzi et al. 1993).
Moreover, in addition to creating new dust grains, 
the planetesimals would further grow into asteroids and also trigger the
formation of planets. In the end, the gaseous parts are 
gradually depleted by the
stellar wind and the debris discs are constructed.

Indeed, high-resolution images of some debris discs show the 
presence of asymmetric density structures or clumps.
Before the extra-solar planets were discovered by the Doppler effect,
these clumpy structures gave indirect evidences of the existence of 
planets. If there were no planets around 
Vega-like stars, it will be much more difficult to explain the 
asymmetric structures of debris discs. 

On the other hand, the astronomers' observational effort 
has led to the rapid progress on the discovery of planets
and there are now more than 200 detected 
extra-solar planetary systems. Many theoretical work on their
dynamical structures have been done 
(Gozdziewski \& Maciejewski 2001, Jiang et al. 2003, 
Ji et al. 2002, and also Ji et al. 2007). 
Moreover, the possible effects of discs on the evolution of planetary systems
are also investigated (Jiang \& Yeh 2004a, 2004b, 2004c).
In fact, some of these systems are associated
with the discs of dust. For example, by a sub-millimeter camera, 
Greaves et al. (1998) detected the dust emission around the nearby star
Epsilon Eridani. This ring of dust is at least 0.01 Earth Mass and the peak
is at 60 AU. It is thus claimed to be the young analog to the Kuiper
Belt in our Solar System. Furthermore, Hatzes et al. (2000) discovered
a planet orbiting Epsilon Eridani by radial velocity measurements,
making the claim in Greaves et al. (1998) even more impressed.

Therefore, the existence of a debris disc implies the presence of 
planetesimals and probably also planets. 
The study of debris discs is thus very interesting and important
because the density structures and evolutionary histories of debris
discs actually provide hints on the evolution of planetesimals and the
formation of planets.  

Since the Vega system gives one of the closest and brightest debris discs, 
many observations have been done and revealed the detail information.
For example, Su et al.(2005) 
showed the 24, 70 and 160 
$\mu m$ images of Vega observed by the Spitzer Space Telescope.
They claimed that the Vega's disc mainly consists of two parts: (i) 
the dust grains, which could be composed of grains of 2 sizes only
or multiple sizes with a power-law size distribution; (ii) a ring
from 86 AU to 200 AU, which is composed of larger asteroidal bodies.
These asteroidal bodies
keep producing new grains, which  
migrate outward and form $1/R$ density profile of the outer disc.
However, 
it is unclear 
whether a self-consistent dynamical model can be constructed
when the assumption in Su et al. (2005)
that the grains of all different sizes are
homogeneously distributed is relaxed.
 
On the other hand, Krivov et al. (2006) used the kinetic method to 
calculate the dust grains' size and spatial distributions expected
in a steady-state debris disc. In addition to the stellar gravity, 
both the radiation pressure and collisional processes are included.
They applied the models on the Vega's debris disc and 
found that the grains in the steady-state disc are dominated
by the smallest ones in bound orbits, which are grains with radius about
10 $\mu$m, and the size distribution has a wave-like pattern.
They thus favor a steady-state disc with a much longer lifetime 
rather than a relatively
short lifetime as in Su et al. (2005).
  
The advantage of Su et al. (2005) is that the results are directly
derived from the observational data and the drawback is that it assumed 
the grains of all different sizes are
homogeneously distributed and was lack of dynamical calculations.
The positive of Krivov et al. (2006) is that the theoretical background 
is stronger and the negative is that there is no connection with the 
observations. Neither is the real chemical composition considered 
in their calculations.

In order to clarify the above issue,  
we here seek
self-consistent dynamical models for the Vega debris disc. 
For those models based on the results in Su et al. (2005), 
we would assume  
the ring region (from 86 AU to 200 AU) keeps generating new grains.
These grains are added into the system and move to where it shall be
according to the equations of motion. The distribution of these grains
will have to produce $1/R$ density profile and also fit the 
observed surface brightness for the outer disc between 200 and 1000 AU.
The classical simple power laws would be used for the size distributions
of these Su-like models (hereafter Type I models).
For those models based on the concept of long-lived grains as in
Krivov et al. (2006), the grains are dominated by the ones 
with size greater than 10 $\mu$m and distributed initially all the way upto
1000 AU. No new dust grains would be added into the system and the size 
distributions are set to be similar with the ones in Krivov et al. (2006).
We would also try to fit the predicted surface brightness of these 
Krivov-like models (hereafter Type II models) with the observational data.

Therefore, the purpose of Type I models is to provide the dynamical support
or to examine the self-consistence of blowing-out picture when the assumption
of grain homogeneity is relaxed.
The reason to do Type II models is to check whether an outer disc dominated 
by the long-lived grains could explain the data.
In this paper, we only discuss the 
structure of the outer disc, i.e. from 200 to 1000 AU, when we compare 
the simulational data with the $1/R$ law or with the observational data.

We present the model and initial conditions in \S 2. In \S 3, the
results of numerical simulations are described. 
Finally, we make conclusions in \S 4.

\section{The Model Construction} 

In our model, the dust grains' motion is governed by the gravity and radiation
pressure from the central star. 
Through the calculations of the orbital evolution of these dust grains,
we will be able to determine 
the density distribution and also the surface brightness of the debris disc 
at any particular time.


For our equations of motion, the unit of mass is $M_{\odot}$,  
the unit of length is AU, and the unit of time is year. 
Thus, the gravitational constant
$G=6.672\times 10^{-11} ({\rm m^3/kg\,sec^2})
=38.925({\rm (AU)^3/M_{\sun} year^2})$, and the light speed 
$c=3\times 10^8{\rm  (m/sec)}=6.3\times 10^4 ({\rm AU/year})$.
All simulations start at $t_0\equiv 0$ and
end at $t_{\rm end}\equiv 11000$.


All dust grains are assumed to be in a 
two dimensional plane, governed by the gravity and radiation pressure
from the central star. 
For any given time, the central star is fixed at the origin and 
the dust grain's equations of motion are 
as in Moro-Martin \& Malhotra (2002). 
Grains in both types of models are assumed to move on circular orbits
initially.

The non-gravitational 
influence on the grain orbital evolution is completely determined 
by the parameter $\beta$. When the chemical composition of dust grains
is chosen, the values of $\beta$ mainly 
depend on the grain sizes. Thus, different types of size distribution
would imply different orbital evolution of grains and the  
density distribution of the debris disc.


It is unclear what kind of size distribution shall be assigned to 
dust grains. Assuming there are enough collisions for 
grains to fragment and combine, it shall follow the classical standard
power law with index -3.5.
This power-law size distribution would be used in Type I models, i.e.
Model Ia and Ib. 

On the other hand, 
Krivov et al. (2006) suggested that
the size distribution at steady-state shall be wavy
for the grain larger than blowing-out sizes.
Thus, a wavy size distribution as in Krivov et al.(2006)
would be used in Type II models, i.e. Model IIa and IIb.

For Model Ia (with C400 grains) and Ib (with ${\rm MgFeSiO_4}$ grains), 
the dust particles are supposed to be produced through the collisions of 
asteroidal bodies in the ring region, between 86 and 200 AU,
the initial positions of  
the dust grains are thus randomly placed in this region.  
From 86 to 100 AU, the surface number density is set to be a constant.
At $R=100\AU$ the surface number density starts to 
decrease as $1/R^2$ until $200\AU$.
After that, we add new grains with the above distribution 
at $t=200 \times i$, $i=1, 2, 3,...,54$ into the system. 


For Model IIa (with C400 grains) 
and Model IIb (with ${\rm MgFeSiO_4}$ grains), 
the particles are uniformly placed in 
the region between $86{\AU}$ and $100{\AU}$. At  $R=100\AU$ the surface 
number density starts to decrease as $1/R^2$ until $200 \AU$. 
At  $R=200\AU$, the surface 
number density starts to decrease as $1/R$ until $1000 \AU$.
We would not add any new grains during the simulations.


\begin{figure} 
 \centerline{{\epsfxsize=10cm\epsffile{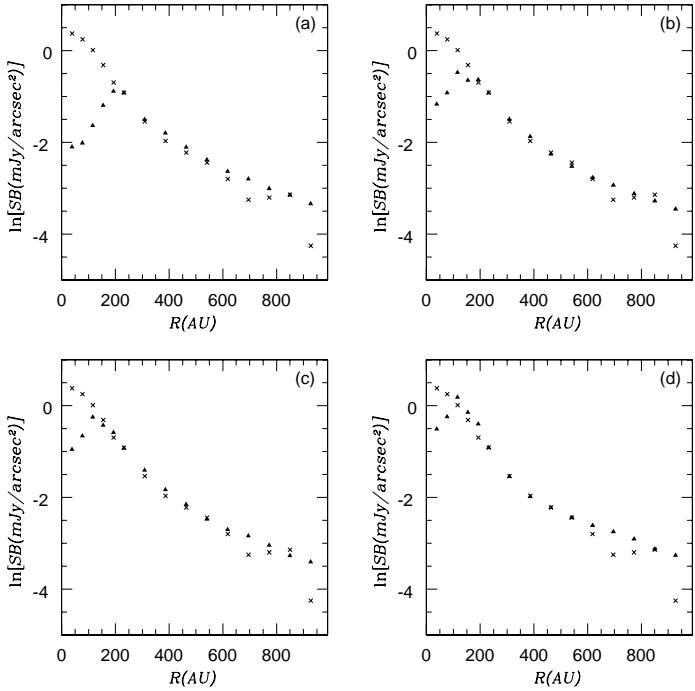}}}
 \caption[]{
The surface brightness of the Vega debris disc
at the end of simulations, $t=11000$:
(a) Model Ia, (b) Model Ib, 
(c) Model IIa, (d) Model IIb.
In all panels, 
the crosses are the Spitzer data, and the triangles are 
the convolved simulational data with best scaling factors $c_{sb}$.
}
\end{figure}

\section{Results}

Based on the above arrangements, 
the simulations were performed and the final profiles of the surface mass
densities in all models are obtained. We found that
the density profile in Model Ia can be well fitted by a $1/R$ law.
This is because that, in Model Ia, the debris disc is dominated by 
smaller grains
and the continuous blowing out of dust grains forms 
a steady-state flow, which naturally gives
a $1/R$ profile. 

The final density profile in Model Ib is a smooth decaying 
curve but the slope is deeper than $1/R$, due to the accumulation
of larger grains. For Type II models, the final density profile in Model IIa 
presents 
a shock transition around $R=700$ AU. Similarly, there is also a shock
transition at 650 AU in Model IIb. This confirms that the grains of 
Type II models are long-lived and cannot move toward the boundary 
of the system.

The corresponding surface brightness at the wavelength $\lambda=24 \mu m$ 
of the above four panels are plotted in Fig. 1. 
We choose $\lambda=24 \mu m$ due to the smaller beam size.
The crosses are for the observed Spitzer data and the triangles are
for the convolved simulational data.
The simulational results can reasonably 
fit the Spitzer data in all models
for the region between 200 and 1000 AU.

Both the total mass at the beginning, $t=t_0$, and the end, $t=t_{\rm end}$,
of simulations are listed in Table 1.
\begin{center}
{\bf Table 1} The Scaling Factors $c_{sb}$ and the Representing Mass
\begin{tabular}{|c|c|c|c|}
\hline 
Model & $c_{sb}$ & $M_{\rm tot}(t_0)$ ($M_\oplus$) & 
$M_{\rm tot}(t_{\rm end})$ ($M_\oplus$)\\\hline
Ia & $4.33\times 10^{29}$ & $1.79\times 10^{-4}$ & $9.48\times 10^{-4}$\\\hline
Ib & $2.36\times 10^{30}$ & $1.43\times 10^{-3}$ & $2.13\times 10^{-2}$\\\hline
IIa & $4.99\times 10^{28}$&$4.48\times 10^{-3}$&$3.08\times 10^{-3}$\\\hline
IIb &$4.26\times 10^{29}$&$5.58\times 10^{-2}$&$5.54\times 10^{-2}$ \\ \hline
\end{tabular}
\end{center}
The initial total mass of dust grains in Model Ia is $1.79\times 10^{-4}$
$M_\oplus$. The same amount of new grains is added every 200 years,
which is equivalent to a dust production rate about $3 \times 10^{14}$ g/s, 
so that the total mass of grains within 1000 AU increases and becomes
$9.48\times 10^{-4}$$M_\oplus$ at $t=t_{\rm end}$.

The total mass of disc grains and the dust production rate of Model Ia are
compatible with the values in Su et al. (2005). Moreover, 
the simulational surface brightness 
of this model could fit the Spitzer data.
Thus, the general picture of 
blowing-out model as claimed in Su et al. (2005) is dynamically valid
and there is no need to assume large grains to be as far as 1000 AU
by arguing that the grains could be very non-spherical. 

On the other hand,
the initial total mass of dust grains in Model Ib is $1.43\times 10^{-3}$
$M_\oplus$. Similarly, this amount of grains is added every 200 years,
which is equivalent to a dust production rate about $2 \times 10^{15}$ g/s. 
This is higher than the value in Model Ia.
Many larger  
grains get trapped within 1000 AU, so that the total mass of grains within 1000 AU increases and becomes
$2.13\times 10^{-2}$$M_\oplus$ at $t=t_{\rm end}$, 
which is about one order larger than the values in Su et al. (2005).

For Type II models,
the total mass of grains within 1000 AU at $t=t_{\rm end}$ 
is very close to the corresponding initial value.
This is because most grains stay within 1000 AU all the time and 
only a small number of them escape.



\section{Concluding Remarks}

In this paper, we investigate the possibilities of the construction of
self-consistent dynamical 
models of the Vega debris disc. We relax the assumption of homogeneous 
distributed grains in the disc and allow the grains to freely move to 
anywhere following the radiation-dynamical equations. The grain's 
orbital evolution is thus calculated 
and the density distribution is then determined.
In order to test different self-consistent models, two chemical 
compositions of grains are considered, and 
the power-law and wavy size distributions are employed for 
short-lived (Type I) and long-lived (Type II) pictures, respectively,

Our results show that the self-consistent dynamical model with a $1/R$
disc density profile
can be constructed when the grains have 
a power-law size distribution.
Moreover, both types of models (dominated by short-lived and 
long-lived grains) 
are consistent with the observational data. 




\label{lastpage}


\end{document}